\documentclass{llncs}

\usepackage{graphicx}
\usepackage[UKenglish]{babel}
\usepackage{url}
\usepackage{textcomp}
\urldef{\mailsa}\path|fmartinfdez@gmail.com|  
\urldef{\mailsb}\path|pcaballe@ull.es| 

\begin{document}

\title{Analysis of the New Standard Hash Function}

\titlerunning{Analysis of the New Standard Hash Function SHA-3}
\author{F. Mart\'in-Fern\'andez\and  P. Caballero-Gil}
\institute{Department of Statistics, Operations Research and Computing\\University of La Laguna. SPAIN\\
\mailsa \ \
\mailsb\\}

\maketitle

\begin{abstract}
On 2$^{nd}$ October 2012 the NIST (National Institute of Standards and Technology) in the United States of America announced the new hashing algorithm which will be adopted as standard from now on. Among a total of 73 candidates, the winner was Keccak, designed by a group of cryptographers from Belgium and Italy. The public selection of a new standard of cryptographic hash function SHA (Secure Hash Algorithm) took five years. Its object is to generate a hash a fixed size from a pattern with arbitrary length. The first selection on behalf of NIST on a standard of this family took place in 1993 when SHA-1 was chosen, which later on was replaced by SHA-2.

This paper is focused on the analysis both from the point of view of security and the implementation of the Keccak function, which is the base of the new SHA-3 standard. In particular, an implementation in the mobile platform Android is presented here, providing the first known external library in this mobile operating system so that any developer could use the new standard hashing. Finally, the new standard in applications in the Internet of Things is analysed.
\end{abstract}

\section{Introduction}

A major paradigm of computing is the handling of large volumes of data. The runtime algorithms most affected by the size of the information as well as manage communications speeds, etc. So it seems obvious that reducing the size of the data we work, usually will increase the speed of our calculations and our communications. One technique to address this problem is called hash functions. A hash function is any algorithm or subroutine that maps large data sets of variable length to smaller data sets of a fixed length, serving a number of properties to ensure data integrity.

Many applications of hash functions are related to the field of cryptography (ciphers, cryptographic accumulators, digital signature, authentication cryptographic protocols, ...). Cryptography is a branch of mathematics that provides tools to achieve security in information systems. Interesting hash functions in the area of cryptography are characterized by complete a series of properties that allow utilities that use cryptographic be resistant against attacks that attempt to compromise the security of the system. A hash functions that satisfy these properties are called cryptographic hash functions. The, cryptographic, hash value, such that an, accidental or intentional, change to the data will, with very high probability, change the hash value. The input of the cryptographic hash function, data to be encoded, is often called the 'message' and the output of the sryptographic hash function sometimes is called the 'digest'. One of the cryptographic hash functions more recognized worldwide, are family owned Secure Hash Algorithm (SHA).

The Secure Hash Algorithm is a family of cryptographic hash functions published by the National Institute of Standards and Technology (NIST) as a U.S. Federal Information Processing Standard (FIPS). In 1993 there was the first publication of a member of this family under the name of SHA. It was withdrawn shortly after publication due to an undisclosed 'significant flaw' and replaced by the slightly revised version SHA-1. SHA-1 was published in 1995 because of these weaknesses of the firstborn of the family. SHA-1 is a 160-bit hash function which resembles the earlier MD5 algorithm. This was designed by the National Security Agency (NSA) to be part of the Digital Signature Algorithm. Cryptographic weaknesses were discovered in SHA-1, and the standard is no longer approved for most cryptographic uses after 2010. Therefore, currently recommend using SHA-2, which is a family of two similar hash functions, with different block sizes, known as SHA-256 and SHA-512. They differ in the word size; SHA-256 uses 32-bit words where SHA-512 uses 64-bit words. There are also truncated versions of each standardized, known as SHA-224 and SHA-384. These were also designed by the NSA. Although SHA-2 is still safe so far, in 2007 NIST published the terms of a public contest to choose the future successor of SHA-2. On 2$^{nd}$ October 2012 the NIST (National Institute of Standards and Technology) in the United States of America announced the new hashing algorithm which will be adopted as standard from now on. Among a total of 73 candidates, the winner was Keccak, designed by a group of cryptographers from Belgium and Italy \cite{SHA3}. The public selection of a new standard of cryptographic hash function SHA (Secure Hash Algorithm) took five years. Its object is to generate a hash a fixed size from a pattern with arbitrary length. 

This paper is focused on the analysis both from the point of view of security and the implementation of the Keccak function, which is the base of the new SHA-3 standard. In particular, an implementation in the mobile platform Android is presented here, providing the first known external library in this mobile operating system so that any developer could use the new standard hashing. Finally, the new standard in applications in the Internet of Things is analysed.

As Section 2 will be explained in more detail this new cryptographic hash function to analyze their security and performance in section 3. Section 4 will detail the software implementation carried out by ourselves. Continue enumerating, in Section 5, applications in the Internet of Things in which can be useful and necessary to introduce the new hashing algorithm standard. Finally give the conclusions and future work to which we have come after the completion of this work.

\section{The new standard hash function}

As mentioned in the introduction, the new SHA family member, to be known as SHA-3, is called Keccak. Keccak is a cryptographic hash function which uses a Sponge Construction \cite{Keccak}, which in a cryptographic context is an operating mode on the base of a fixed length transformation and padding rule. Particularly it is a generalization of hash function which have a fixed output length and the stream ciphers that have a fixed input length. 

The Sponge Construction can be used to implement various cryptographic utilities like hash, pseudo-random number generators, key generations, encryption, Message Authentication Codes (MAC) and authenticated encryption. The Sponge Function operates on a $b=r+c$ bits state. The input array of bits is padded with a reversible padding rule and is split in blocks of $r$ bits. Then, the $b$ bits of state which will be both the input and output of the hash function which will be use in the Sponge Construction, will be initialized to '0'. Once the parameters to be use in the Sponge Function are initialized, the application will be proceeded in two pattern stages in which it will be divided. The first phase is called \textit{absorbing} and it consist of applying the hash function as many times as $r$ blocks are split in the $M$ input. The hash function will be receive in each iteration, as input, the $XOR$ operation of the $r$ block from that iteration and the state which is operating on. This state will be changing by the hash function. The second stage, called \textit{squeezing}, will start when the \textit{absorbing} phase is over and it has the object of getting the $n$ bits output hash. At this stage the first $r$ bits of state resulting from the application of the first stage are already output bits. If more bits were needed to complete the $n$ bits output, the hash function will be applied on the state as many time as necessary.

\begin{figure}
	\centering
		\includegraphics[height=4.2cm]{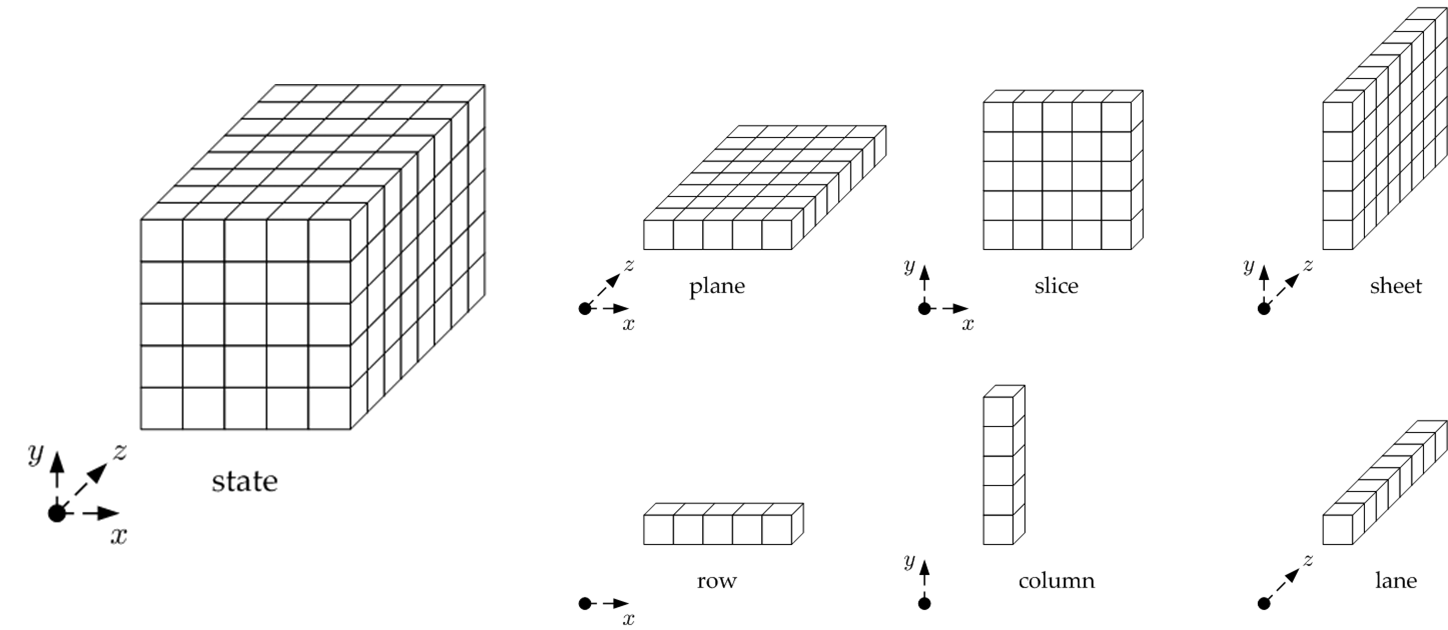}
	\caption{State of SHA-3}
	\label{fig:stateSHA3}
\end{figure}

In the SHA-3 Keccak's version a state represented by $5x5$ matrix of 64 bits lanes will be used (see Figure~\ref{fig:stateSHA3}). The length of output $n$ can adopt four different values: $224, 256, 386, 512$, which will condition the rest of parameters: $r, c and b$. Before starting with the \textit{absorbing} stage, to make sure that the message could be split in $r$ bit blocks, is first padded with a minimum of $10\cdot1$ bits pattern. This pattern consist of '1' bit, zero or more '0' bits (maximum $r-1$) and a final '1' bit. In Keccak, the function is iterated $12+2$\l times, being in SHA-3 \l=6, which is the maximum value that Keccak can adopt. The basic transformation of Keccak involves 5 steps:
\begin{itemize} 
\item Theta ($\theta$), which is $XOR$ operation of each bit in the state with $XOR$ operation of the value of one column that in the same slice but in adjacent column and again $XOR$ operation with another adjacent column that is not in the same slice but in an adjacent sheet.
\item Rho ($\rho$), in which the bits are shifted in their lane by a given number of fixed transformation bits.
\item Pi ($\pi$), which is a row permutation of columns.
\item Chi ($\chi$), the only non lineal operation where an $XOR$ operation of a particular column is performed with the $AND$ operation of negation of the adjacent column and the column next to it.
\item Iota ($\iota$), which is the $XOR$ operation of a round constant into one lane of the state.
\end{itemize}

\section{Security and permormance}

One of the major requirements of any cryptographic hash function that must be taken is to know how much secure is. The analysis Keccak it faced \cite{CompetitionNIST}, in proofs of security relating to the hash functions structure, concluded that Keccak has indifferentiability (the hash function behaves like a random oracle, which is a perfectly random function that can be evaluated quickly) proofs, which guarantee that the hash function resists generic attacks up to at least the complexity of a brute-force collision attack (it consists in finding two arbitrary inputs that produce the same hash), assuming that the underlying permutation primitive is ideal. Also, Keccak is proven to have near-optimal security regarding the core properties of collision, preimage (preimage resistance means that it is hard to find a pre-image from its known hash), and  second-preimage resistance (Second-preimage resistance means that, given an input, it should be difficult to find another input such that the hash digests of both inputs coincide).

Another of the tests he underwent Keccak was that of cryptanalysis on the hash function or its components that relate directly to the core security properties of a hash function. The cryptanalysis results clarified that an attack on six rounds of a ten-round hash function would give a 40\% Security Margin (it is the fraction of the hash function that has not been successfully attacked), but Keccak has 79\% of its hash function still unbroken.

Another test that happened was that related to side channel analysis. A side channel attack is any attack based on information gained from the physical implementation. It can be used to reveal some secret processed by a hash function, such as the key in a HMAC computation. Keccak does not have any non-invertible steps, so an attacker who learns the entire intermediate state for any HMAC-Keccak computation can use this to determine the original key used for the HMAC computation, and can forge arbitrary messages for this key. However, since Keccak uses logical operations, it is easy  to protect it against side channel attacks.

All these results were a factor in the NIST chose to Keccak as new standard for SHA-3. One of the handicap is its hashing speed Keccak in software. For example if used SHA-3 with 512-bit output, the hashing speed, software level, is slightly slow. Under these conditions and for long messages (greater than 4096 bytes) should take about 22.4 cycles/byte, on the reference processor proposed by NIST that it is a Sandy Bridge Desktop Processor (Intel Core i7-2600k) with Current Vector Unit. SHA2-512 take about 14 cycles/byte. But, on the other hand shows that SHA-3 resists any attack up to 2$^{512}$ operations. It would take 4.2*$10^{128}$ years to evaluate the permutation 2$^{512}$ times with one billion computers, each performing one billion evaluations of Keccak-f per second (3*10$^{118}$ times the estimated age of the universe). Note that just counting from 1 to 2$^{512}$ with an irreversible computer working at 2.735 \textdegree K would take at least 3.5*10$^{131}$ joules, which is the total energy output of the Sun during 2.9*10$^{97}$ years.

Due to the nature of Keccak, its implementation in software is conveniently implemented as 64-bit logic operations, rotations, loads and stores. However, general-purpose computers cannot exploit most of the parallelism latent in the algorithm, and a full hardware implementation of Keccak is naturally highly parallelizable, resulting in a very good throughput/area ratio, typically about twice or more of the throughput/area ratio of a full-round SHA-2 implementation.

\section{Our software implementation}

For this work, we developed a Application Programming Interface (API) in the Java programming language, specifically in version 6. This API is fully developed and functional, with full support for the Android mobile platform. The API design is intended for didactic use and it has a hexadecimal as input and returns another hexadecimal. We did not find any published Java code of SHA-3. As a guide for implementation, we used the official development optimized in C language and the official pseudocode. Our implementation has the same structure as the official one, but does not use the same data types.

The API consists of different class, but basically 3 class are importants. The class named SHA-3 calls to the other two classes that are: on one hand the class to represent the sponge function and the other representing the cryptographic hash function of Keccak. The sponge function class has the absorbing function, which is responsible for calling the class that has the steps of the cryptographic hash function, and the function of squeezing. The following code segment has the absorbing function that it calls the cryptographic hash function as many times as $r$ blocks are split in the message input.

\begin{verbatim}
public static void absorbing(State state, byte[] mess, Constants const){
   byte []padKeccak = Padding.apply(mess, const.r);
   int iPadKeccak=0;
   byte[] r = null;
   while (paddingKeccak.length > iPadKeccak){
      r = Utilities.getRInArrayByte(padKeccak, iPadKeccak, const.r);
      BitsOperations.rStateXORrMess(state, r);
      Keccak.hashFunction(state, const);
      iPadKeccak+=const.r;
   }
}
\end{verbatim}

The following code segment has the squeezing function of the sponge function class. The squeezing phase, for Keccak in SHA-3, no mire than squeezing. Simply return to the output of the last iteration of the absorbing phase.

\begin{verbatim}
public static String squeezing(State state, Constants const){
   String ret = "";
   int outputLength=const.n;
   int outputCounter=0;
   String laneHex;
   for(int y=0; y<State.Y; y++){
	    for(int x=0; x<State.X; x++){
         laneHex = Utilities.laneToInverseHex(state.getLane(x, y));
         for (int i=0; i<laneHex.length(); i++){
            ret+=laneHex.charAt(i);
            outputCounter+=4;
            if (outputCounter == outputLength)
               return ret;
         }
      }
   }
   return ret;
}
\end{verbatim}

The cryptographic hash function iterates the 5 steps described above in section 2, a total of 24 times. In our implementation we have joined Rho and Pi steps as did the original pseudocode Keccak authors, as we see the following code:

\begin{verbatim}
public static void hashFunction(State state, Constants const){
   for (int i=0; i<Const.iterationNumbers; i++){
      // Step 1
      theta(state);
      // Step 2 and 3
      rhoYPi(state);
      // Step 4
      ji(state);
      // Step 5
      iota(state, i);
   }
}
\end{verbatim}

Finally, each of the steps the state operates as explained in section 2.

\begin{figure}
	\centering
		\includegraphics[height=4.2cm]{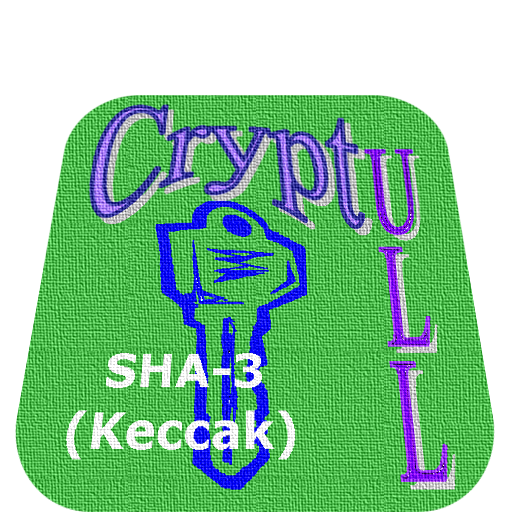}
	\caption{Android application icon in Google Play}
	\label{fig:icon}
\end{figure}

Once the API was designed, to test its performance we created a very simple Android application that generates SHA-3 hash. The application is uploaded from October 2012 on Google Play named SHA-3 Generator (see as is the logo of the application on Google Play in Figure~\ref{fig:icon}). People are starting to use the application and comment on it, so that begins to exist feedback will be used to improve future versions and make the requirements proposed by users.

\section{Applications in the Internet of Things}

One area of particular interest to our research group is the so-called Internet of Things. Internet of Things pretends to merge the real world with the virtual world through interconnection of any everyday thing.

We propose that SHA-3 may be used to generate:

\begin{itemize}
	\item The digest of passwords or files to be verified.
	\item Challenges and/or responses in a challenge-response authentication protocol.
	\item The digest of messages to be signed with a digital signature scheme based on public keys.
	\item A message Authentication Code (MAC) based on secret keys. Unlike SHA-1 and SHA-2, SHA-3 does not have the length-extension weakness in which an attacker, given only H(M) for some unknown message M, can append additional own blocks. Hence, SHA-3 does not need the HMAC nested construction where the key is used twice in order to use the hash to build a MAC. Instead, MAC computation can be performed by simply prepending the message with the key.
\end{itemize}

Thus, it may be used in comunnications between objects of the Internet of Things.

\section{Conclusions and future works}

In this paper, we have analyzed the new standard hashing Sha family NIST announced last October. After an official competition that began in 2007, the winner was Keccak one cryptographic hash function that uses the so-called sponge construction. Considered SHA-3 is secure, at least for now and it will be the replacement of SHA-2, although the latter is recommending continued use of time. Due to SHA-3 Design, Implementation hardware is better than a pure software approach. Thus, the performance of software Implementations must be optimized. We have also proposed a Java API, which is fully functional on Android. In this platform, we have developed an application, for teaching purposes, which generates SHA-3 hash. This application is currently the only one in the Google Play makes use of the new standard hashing.

In addition, this paper leaves open several lines of future work. We will improve our implementation of SHA-3 by optimizing the code written in Java language or rewriting the code using C language for Android. Also, we would like to enhance the documentation level by adding more detail. We could keep looking for the weaknesses of SHA-3. And specially we will develop applications of SHA-3 in the Internet of Things.

\end{document}